\documentclass{vgtc}                          

\ifpdf
  \pdfoutput=1\relax                   
  \usepackage{graphicx}                
  \DeclareGraphicsExtensions{.pdf,.png,.jpg,.jpeg} 
\else
  \usepackage{graphicx}                
  \DeclareGraphicsExtensions{.eps}     
\fi%

\graphicspath{{figures/}{pictures/}{images/}{./}} 
\usepackage{microtype}                 
\PassOptionsToPackage{warn}{textcomp}  
\usepackage{textcomp}                  
\usepackage{mathptmx}                  
\usepackage{times}                     
\usepackage{cite}                      
\usepackage{tabu}                      
\usepackage{booktabs}   
\usepackage{caption} 
\usepackage{color}

\newlength{\dummylen}

\onlineid{0}

\vgtccategory{Research}

\vgtcinsertpkg

\title{Recognizing Handwritten Source Code}

\author{Qiyu Zhi\thanks{e-mail: qzhi@nd.edu} %
\and Ronald Metoyer\thanks{e-mail:rmetoyer@nd.edu}}
\affiliation{\scriptsize University of Notre Dame}

\abstract{
Supporting programming on touchscreen devices requires effective text input and editing methods.  Unfortunately, the virtual keyboard can be inefficient and uses valuable screen space on already small devices.  Recent advances in stylus input make handwriting a potentially viable text input solution for programming on touchscreen devices.  
The primary barrier, however, is that handwriting
recognition systems are built to take advantage of the rules of natural language, not those of a programming language. In this paper, we explore this particular problem of handwriting recognition for source code.  
We collect and make publicly available a dataset of handwritten \textit{Python} code samples from 15 participants and we characterize the typical recognition errors for this handwritten \textit{Python} source code when using a state-of-the-art handwriting recognition tool.  We present an approach to improve the recognition accuracy by augmenting a handwriting recognizer with the programming language grammar rules. Our experiment on the collected dataset shows an 8.6\% word error rate and a 3.6\% character error rate which outperforms standard handwriting recognition systems and compares favorably to typing source code on virtual keyboards.
} 

\CCScatlist{ 
  \CCScat{H.1.2.}{User/Machine Systems}{Human information processing}{};
  \CCScat{H.5.2.}{User Interfaces}{Input devices and strategies (e.g., mouse, touchscreen)}{}
}

\begin{document}



\maketitle

\section{Introduction} 
With the rapid technology shift in current computing devices, high-quality low-cost mobile devices such as tablets and smartphones are being increasingly used in everyday activities. Many tasks that previously required a PC are now feasible on mobile devices. For example, tablets are typically equipped with powerful batteries, advanced graphic processors, high-resolution screens and fast processors, making writing and compiling code on them completely plausible. TouchDevelop, for example, is a novel programming environment, language and code editor for mobile devices  \cite{tillmann2011touchdevelop}. Furthermore, Tillmann et al. predict that programming on mobile devices will be widely used for teaching programming \cite{tillmann2012future}.  However, mobile devices are also inherently restricted by their limitations such as small screens and the clumsy virtual keyboard.  Entering and editing large amounts of text for programming tasks can quickly become difficult and time consuming with these virtual keyboards because they are notoriously difficult to use when compared to a physical keyboard and they consume valuable screen space\cite{raab2013refactorpad}.


While keyboards have been the primary input device for entering computer programs since the computer was invented\cite{gordon2013improving}, this predominant mechanism is not ideal for all programming situations. For example, software developers that suffer from repetitive strain injuries (RSI) and related disabilities may find typing on a keyboard difficult or impossible \cite{begelprogramming}.
Instead, handwriting with a stylus may be a preferred input mechanism for some of these users \cite{mankoff1998cirrin}.
In addition, some physical configurations (e.g. seated on a plane) may simply be more suited to the writing posture than a typing posture for many users.

Handwriting has also been shown to have potential cognitive benefits\cite{alonso2015metacognition}.  In particular, Mueller and Oppenheimer found that students who took longhand notes performed better on conceptual questions than those that typed notes on a laptop \cite{mueller2014pen}.  Given these findings, and that fact that many programmers write pseudocode by hand before typing, it is reasonable to consider that handwriting may provide cognitive benefits for programming, especially on mobile devices. 
Furthermore, recent advances in pen-based input and handwriting recognition technology are quickly making handwriting a viable alternative to typing.



In this paper, we explore the use of handwriting as a means for source code text input.  There are two ways to approach this problem.  One alternative is to develop or modify a handwriting recognition engine to take source code directly into account.  Given that source code often includes English language words, another alternative is to leverage the capabilities of an existing English language handwriting recognition engine. We explore this latter option.  First, we collect and present a publicly available dataset of handwritten \textit{Python} source code for use in handwriting recognition research.  Second, we explore the use of the state-of-the-art recognition system, MyScript \cite{myscript} for recognizing \textit{Python} source code.  We characterize the errors made by the MyScript engine and present a method for post-processing the engine's results to improve recognition performance on handwritten \textit{Python} source code.   

After presenting related work and necessary background information, we describe our data collection process and the resulting publicly available dataset. We then describe the performance of MyScript on recognizing the handwritten source code and present our algorithm for leveraging the MyScript engine to produce improved results.  We discuss those results in Section \ref{results} and conclude with avenues of future work.




\section{Background and Related Work}


Many alternatives to typed source code have been considered, typically in the context of making programming more accessible.   Most of those appoaches fall in the realm of speech-based programming \cite{desilets2006voicecode, gordon2013improving}, however speech is not always an acceptable solution, especially in quiet environments or with applications that require privacy.
Given the recent advances in pen-based input, handwriting is a potentially viable alternative.  The pendragon supports people who are unable to use
a keyboard and seeks to find new interaction techniques for input which may improve communication speed \cite{Pendragon}. Mankoff et al.also suggest that word prediction, sentence completion and the syntax of programming languages could be used for handwriting source code \cite{mankoff1998cirrin}. 
Most closely related to our work is a programming IDE integrated with a handwriting area in which the handwritten code is recognized by an enhanced handwriting recognition system \cite{frye2008pdp}.  This work, however, does not present an evaluation of the recognition engine or guidance for how to improve general handwriting recognition engines for application to source code recognition.
In this paper, we focus on the recognition step of handwritten source code as the most important step for developing an effective handwriting interface for source code input and editing.

Research in handwriting recognition has a long history dating back to the 1960s~\cite{tappert1990state}. Hidden Markov Model (HMM) based handwriting recognition~\cite{hu1996hmm,kundu1988recognition,nag1986script} is one of the 
most widely used approaches while neural networks are gaining in popularity~\cite{jaeger2001online}. Some approaches also leverage additional constraints for recognizing handwriting in specific domains such as postal addresses \cite{srihari1993recognition, srihari1993interpretation} and banking checks \cite{gorski1999a2ia, agarwal1997bank}. These handwriting recognition systems are developed to take advantage of the English language \cite{van2003using}, which is intrinsically different from source code. For instance, variable names are often created from
concatenated words (e.g. camelCase or underscore naming), which poses a problem for the traditional handwriting
recognition system as it expects spaces to appear between words
contained within its dictionary.   We do not aim to contribute to the extensive literature in handwriting recognition, but rather, we intend to examine how we can leverage this existing work for application to handwritten source code recognition interfaces.

\section{Data Collection}

Our first contribution in this paper is a data collection study designed to generate a sample set of handwritten source code for research purposes.  The first question to consider is what programming language to study.  We decided to collect handwritten \textit{Python} source code because of the current popularity of \textit{Python}\footnote{\url{http://www.tiobe.com/tiobe-index/}} and its projected growth rate 
\cite{radenski2006python}. 
We chose a ``copying task'', where three code samples are provided for every participant to copy on the tablet using the stylus. While we understand that a ``copying task'' may be cognitively quite different from other writing tasks that require synthesis, we sought to eliminate sources of cognitive load that could impact timing as well as writing quality for the purposes of this data collection task. The three shared code samples allow for comparison across participants. To broaden our dataset of unique handwritten source code samples, we also randomly selected a fourth source code sample function (per participant) that was unique to that participant. In this section, we describe our data collection process and the resulting database of handwritten \textit{Python} samples.



\subsection{Participants}
We recruited 15 participants (9 females) from the University of Notre Dame for our study. Thirteen of the participants were computer science majors and all participants had at least two semesters of programming experience. Their ages ranged from 19 to 29 (mean = 22.3). Two participants were left-handed. Eight participants had used a pen/stylus for handwriting on a touchscreen device and only one participant had used a tablet for inputting source code (via the virtual keyboard). All participants were compensated \$5 for the study which took approximately 30 minutes each.

\subsection{Apparatus and Software}

\begin{figure}[tb]
 \centering 
 \includegraphics[width=\columnwidth]{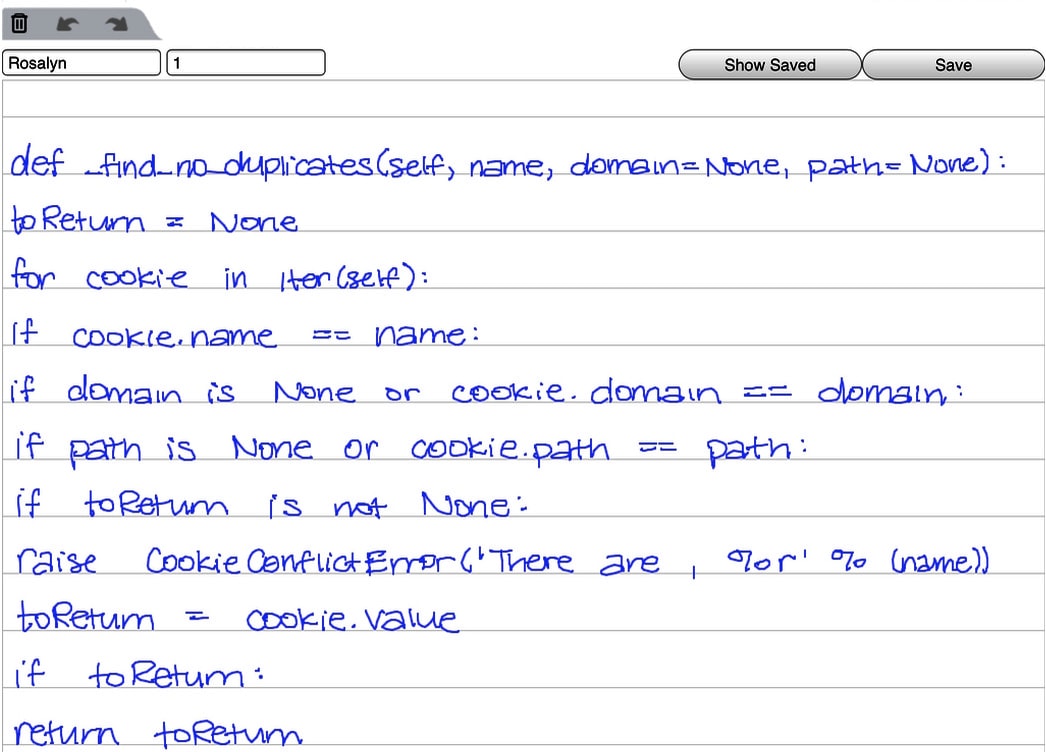}
 \caption{Screen shot of our data collection web application.  Participants entered their name and code sample number into the boxes in the upper left.  They then entered the code sample using the Apple Pencil and selected `Save' when finished.}
 \label{fig:webpage}
\end{figure}

We used a 12.9 inch iPad Pro with a 2732-by-2048 screen resolution at 264 pixels per inch (PPI) and fingerprint-resistant oleophobic coating. Participants used the Apple Pencil as the stylus device. We implemented a web application with a writing area to record user input. This application was responsible for converting touch points of the stylus into handwriting strokes and saving strokes to a JSON file. Each stroke consists of the coordinates of the sampled points and time-stamp information for each coordinate. The writing area in this application measured 795 * 805 pixels with subtle lines on the background to provide guides for the participants (See \autoref{fig:webpage}). We also implemented functions to undo or redo the previous stroke as well as clear the writing area of all strokes. 

\subsection{Representative Source Code Material}
Our goal was to create a database of representative samples of handwritten \textit{Python} source code for use in evaluating the performance of a handwriting recognition system. Because different \textit{Python} samples contain different language elements, there is no single representative corpus\cite{almusaly2015syntax}. Ideally, representative code samples should contain a variety of language constructs and not be restricted to a single project. 

Our process for choosing source code samples is based on that used by McMillan et al. \cite{mcmillan2012exemplar, rodeghero2014improving}. First, we selected six popular \textit{Python} projects on Github. \autoref{table:project} summarizes the details of these projects. We then extracted all functions from the project source code and eliminated comments in order to focus solely on the source code of the samples.  Next, to obtain functions that were sufficiently long to collect a substantial amount of handwriting, but not so long as to require multiple pages of handwriting, we filtered the functions to those with between 9 and 18 lines of source code and those with no lines greater than 60 characters (to eliminate long, wrapping lines).  We also manually filtered out highly repetitive functions, such as a function that includes only assignment statements for variables. The result was 1324 eligible functions. We randomly selected the three shared test code samples from this set for use by all participants and one additional unique code sample to be entered by each participant in the study. Although \textit{Python} syntax considers whitespace, we decided to ignore indentation for the purposes of focusing purely on handwriting recognition. 


\subsection{Procedure}

For every participant, we began our data collection with an informed consent process.  Each participant then filled out a pre-study questionnaire about demographics and experience using touchscreen devices and a stylus. Participants were given a practice task to familiarize them with the process.  For each of the four tasks, participants were given a sheet of paper with the sample typed \textit{Python} source code.   Participants entered their name and code sample number into the web application and then entered the code sample using the Apple Pencil and selected `Save' when finished.  After completing all four input tasks, participants were compensated and the session ended.

\subsection{Data Collection Results}
The final dataset includes stroke data for four code samples for each of 15 participants resulting in a total of 60 handwritten source code samples. So, for each of 3 given source code input samples, we have 15 copies of handwritten source code (for a total of 45 handwritten source code samples). The remaining 15 are handwritten samples of unique input source code examples from each participant. 
The handwritten source code data can be downloaded at \newline \url{http://www.purl.org/recognizinghandwrittencode/data}.
 
\begin{table}
  \centering
  \begin{tabular}{l r r r}
    {Project}
    & {Lines}
      & {Fuctions}
    & {Eligible Functions} \\
    \midrule
    AlphaGo & 1,963 & 151 & 1 \\
    Bittorrent & 7,164 &  570 & 39 \\
    Blender & 265,684  & 12,774 & 1,126 \\
    Instagram & 1,265 & 145 & 8 \\
    Requests & 14,009  & 862 & 84 \\
    Webpy & 10,199 & 1,029 & 66 \\
  \end{tabular}
  \caption{\textit{Python} projects used for selecting code samples}~\label{table:project}
\end{table}

\section{Source Code Recognition Errors}

Current commercial handwriting recognition systems are built to take advantage of the rules of the English language as opposed to that of a programming language, therefore it is not surprising that these systems might perform poorly on source code recognition \cite{frye2008pdp}. There is, however, no previous research that evaluates how well existing state-of-the-art handwriting recognition systems perform on handwritten source code. Here we describe the state-of-the-art handwriting recognition system we employed and characterize the errors based on the dataset we collected.

\subsection{State-of-the-art: Myscript}
Automatic recognition of handwriting is now a mature discipline that has found many commercial uses\cite{plamondon2000online}. MyScript\cite{myscript} is an online handwriting recognition engine that supports more than 80 languages and achieved the best recognition rate in the International Conference on Document Analysis and Recognition competition\cite{el2011line}.
Here we use the MyScript engine as our baseline for comparison and study the typical recognition errors produced when applied to handwritten source code to better understand the complexities introduced by \textit{Python} source code and source code in general.

\subsection{Data Pre-Processing}

In order to use MyScript efficiently and to make a fair comparison between its performance and our algorithm, we apply two simple pre-processing steps to the data. First, we provide MyScript with \textit{Python} specific context through the Subset Knowledge (SK) facility and a custom lexicon. SK is a MyScript feature for telling the recognizer that we only want it to enable recognition of certain characters. For example, for a phone number field, we may want only digits to be recognized.   We created an SK resource in MyScript to allow only characters that can legally appear in \textit{Python} source code.  We also provide the legal \textit{Python} keywords through a user-defined lexicon.

The MyScript Cloud Development Kit (CDK) is an HTTP-based set of services that take handwritten strokes as input and produce potential recognition results as output. To use the CDK in experiments, we must send strokes to the recognizer at some level of granularity (e.g. single character, whole word, whole line, etc).
We chose to break the stroke data into lines assuming developers might write a statement at a time on a single line. To do so, we analyze stroke coordinates and create a new line each time the user moves to a new vertical position. Each line is then sent one at a time to the MyScript CDK.  This simulates a developer writing one programming statement (one line) at a time, pausing at the end of each line.  Alternative pre-processing is possible given the raw stroke data and timestamp information (e.g. sending incomplete lines when a participant pauses).

\subsection{Characterizing errors}
\label{sec:characterizing}
We processed all of the handwritten data as described above to collect baseline recognition results for all handwritten samples in our dataset.  We then set out to understand the types of recognition errors that were present in the final recognized text. We identified three major types of recognition errors: word errors, symbol errors, and space errors. 

Word errors occur when MyScript simply incorrectly recognizes a written word. This is typically due to poor writing and can occur for keywords as well as non-keywords.
For example, when the handwritten word `self' is recognized as `silt', we characterize this as a word error. 
Symbol errors represent incorrect recognition of symbols or non alpha-numeric characters.
For example, an `\_' (underscore) is often recognized as a `-' (dash). 
Finally, a space error results when the system inserts an unexpected space.
For example, when `ConflictError' is recognized as `Conflict Error', we characterize it as a space error. 

Most of the word errors and symbol errors can be attributed to poor writing or cursive writing (characters are written joined together in a flowing manner) which is inherently more difficult for MyScript to recognize than block writing (characters are written separately). Space errors, on the other hand, appear to depend on the language model of the recognizer, which most likely does not include training on CamelCase\footnote{\url{https://en.wikipedia.org/wiki/Camel\_case}} or proper English words separated by dot notation (e.g. student.name).  The result is that MyScript inserts space at these word and dot notation separators.

\begin{figure}[t!]
\centering
\includegraphics[width=3.16in, height = 2.2in]{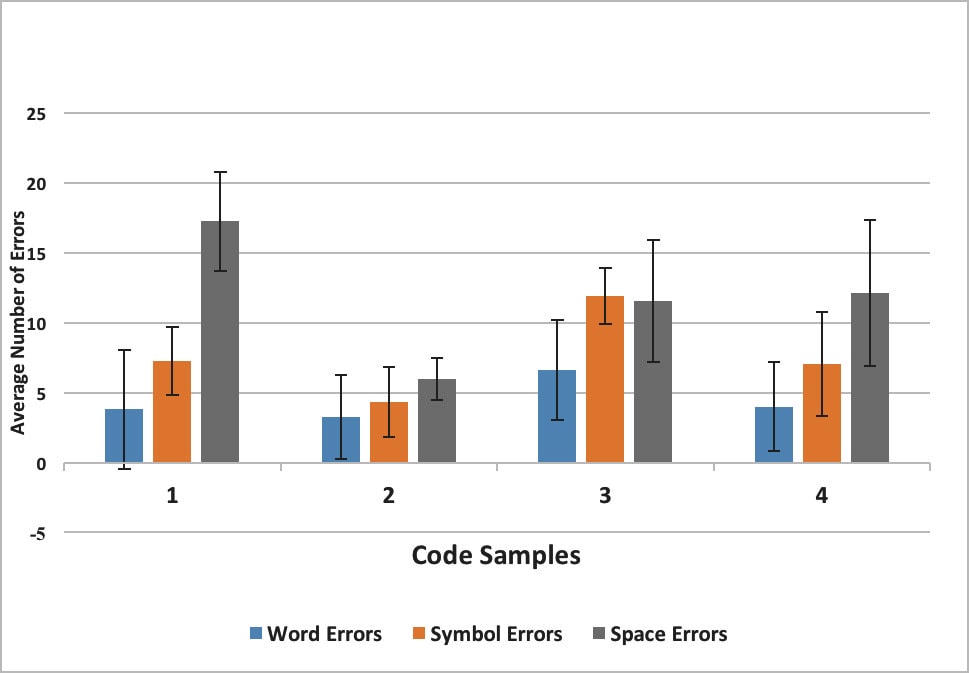}
\caption{Average error numbers of all participants for each code sample from MyScript general handwriting recognition engine}
\label{fig:errors}
\end{figure}

In summary, from the statistical results for each type of error presented in \autoref{fig:errors}, space errors, mainly caused by the internal mechanism of English handwriting recognition system, represent the most prevalent recognition error. In addition, poor writing and the tendency to return an English word for a non-English word in the source code lead to word errors, which also represents a significant portion of all errors. Symbol errors are also a prevalent error type. This makes sense given that MyScript is designed to recognize general words, however, symbols, dot notation, and combinations of symbols and words are typically not present in general text, especially in the way that they are used in source code. For example, the most problematic symbols includes underscore `\_', parentheses `( )' and equal `='.

\section{Handwritten Source Code Recognition Pipeline}

A programming language is governed by grammar rules, which stipulate the positions of keywords and symbols. For example, in \textit{Python}, a \textit{def} sentence must end with a `:'. However, handwritten symbols are often problematic.  For example, colons `:' are sometimes recognized as semicolons `;'. In addition to grammar rules, programming languages are highly repetitive with predictable properties\cite{hindle2012naturalness}. 
Function names and variable names are the most common repetitive words in a single source code project. If a function name appears more than once in the same handwritten code sample, however, it is impossible for users to hand write the \textit{exact} same strokes for this function name, which makes different recognition results of the same handwritten function name a possibility that we must account for.

In this section, we present an approach to improve the recognition rate for handwritten source code by addressing these issues as well as those common errors characterized in Section \ref{sec:characterizing}. We leverage what we know about the predictability and structure of source code to improve recognition results beyond that of the state-of-the-art recognizer.

\begin{figure}[h!]
\centering
\includegraphics[width=0.5\textwidth]{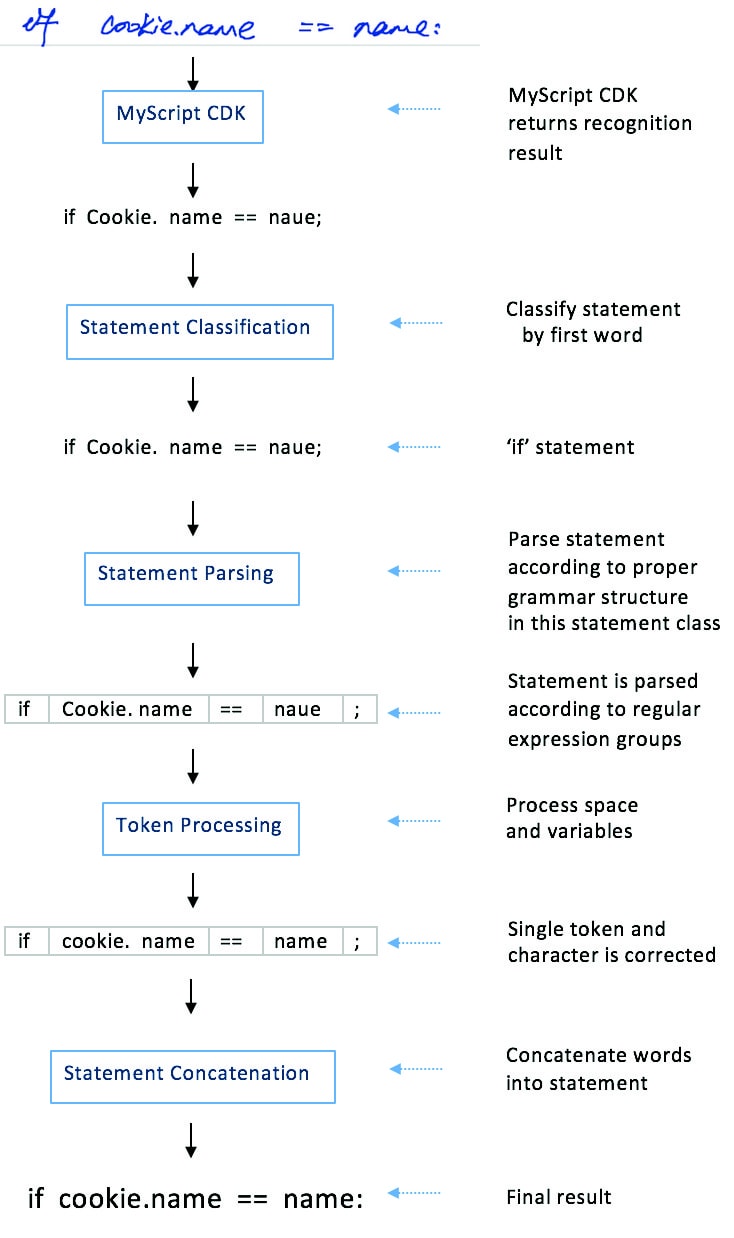}
\caption{Framework for augmenting MyScript to correct for common recognition errors in handwritten source code.}
\label{fig:overview}
\end{figure}

The general premise of our approach is that state-of-the-art engines can produce excellent results given good writing and the absence of symbols and programming practices like camelCase.  Our framework, illustrated in \autoref{fig:overview}, is therefore aimed at analyzing and post-processing the recognition results produced from MyScript to utilize its recognition capabilities but correct for those common errors. This framework can be divided into four parts: statement classification, statement parsing, token processing, and statement concatenation. 
The source code for this post-processing algorithm can be found at \url{http://www.purl.org/recognizinghandwrittencode/code}.

\begin{figure*}[h]
 \center
  \includegraphics[width=2.2in, height = 1.32in]{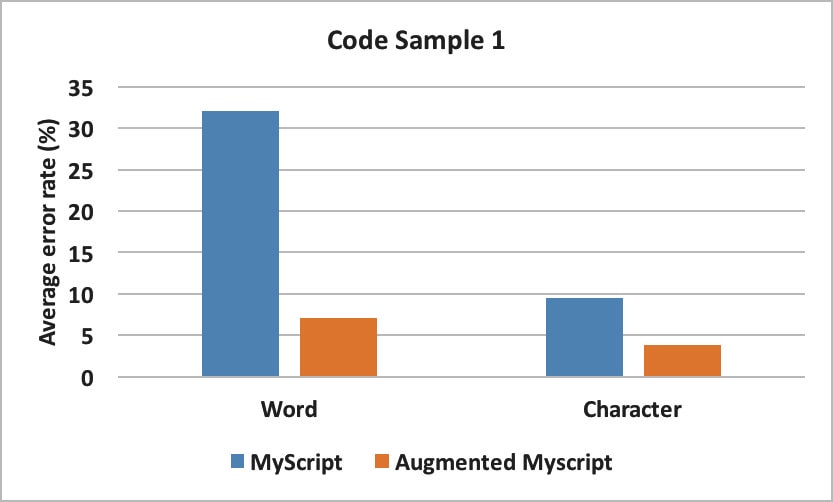}
  \includegraphics[width=2.2in, height = 1.32in]{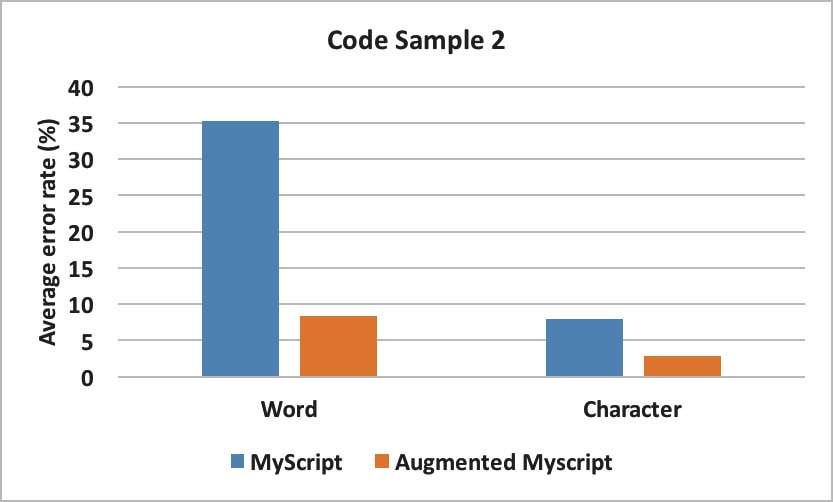}
  \includegraphics[width=2.2in, height = 1.32in]{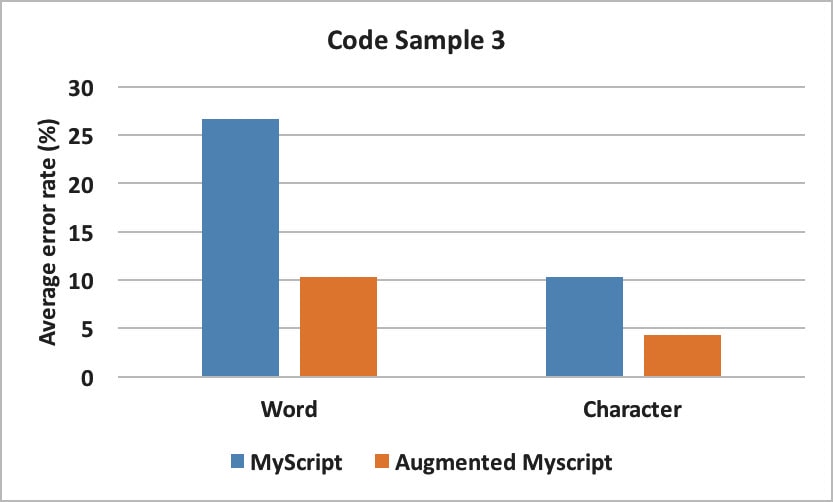}
  \caption{Average recognition error rate of MyScript and our augmented MyScript system for three test code samples}
  \label{result}
\end{figure*}

\subsection{Statement Classification}
As we mentioned before, we process the handwritten source code data considering each statement as a unit. According to the \textit{Python} grammar specification, we can restrict \textit{Python} source code statements into a limited number of classes, each of which has specified structure rules\footnote{\url{https://docs.python.org/2/reference/grammar.html}}. Here we use the first token in the statement as the symbol for classification. For example, a `def' statement starts with `def' and its structure is defined as `def' + `function name' + `(parameters0, parameters1 ...):'. We define 14 classes for \textit{Python} code statements, including an `assignment' statement, which means the first word in this statement is not a keyword but rather a variable name. In \autoref{fig:overview}, the recognition result is classified as an `if' statement. \autoref{table:statement} presents statistics for the various statement classes in the three code samples.

\begin{table}
  \centering
  \begin{tabular}{l r r r}
    {Class}
    & {Frequency}
      & {Class}
    & {Frequency} \\
    \midrule
    def & 3 & except & 1  \\
    if & 7 &  while & 1 \\
    for & 3  & try & 1 \\
    raise & 2 & break & 1 \\
    return & 2  & else & 1 \\
    yield & 2 & assignment & 13 \\
  \end{tabular}
  \caption{Frequency for each statement class in three test code samples}~\label{table:statement}
\end{table}

\subsection{Statement Parsing}
After classifying the statement, we need to break it down into independent parts according to the grammar rules.
Similar to a recursive-descent parser \cite{van1993recursive}, our system consists of a series of functions, each of which is responsible for one class of statement. Each function includes a set of mutually recursive procedures where each such procedure implements one of the productions of the grammar as a regular expression. We implement a top-down LL parser to parse the input from left to right and perform a leftmost derivation \cite{fernau1998regulated} of the statement. As a result, a statement is parsed into a list of single tokens and/or characters. For example, the statement in \autoref{fig:overview} is parsed into five individual tokens. Specifically, `if' is a keyword token; `Cookie. name' is a variable token; `==' is a symbol token; `naue' is a variable token; `;' is the last symbol token.

\subsection{Token Processing}
The previous stage results in a list of single tokens and/or characters that make up the statement.
We assume all non-keywords are properly recognized and add them to the lexicon assuming they are \emph{variable} names.
Then for all non-keywords in each statement that follows, we first compare the token to all the words in the non-keyword lexicon. If a `similar' token already exists in the lexicon, we replace it with the `similar' token in the lexicon. For example, in \autoref{fig:overview}, `naue' is very similar to `name', which is already in the lexicon, so we just replace the token `naue' with `name'. If there is no `similar' token in the lexicon, we accept this token as it is and add it to the lexicon. We calculate similarity using the Levenshtein distance \cite{levenshtein1966binary} with a threshold of 0.7, determined empirically.

\subsection{Statement concatenation}
After processing all tokens, we remove all extra spaces in any single token, then concatenate each token with a single space between them to reconstruct the final statement.
Additionally, we ensure that the last recognized character of a statement is a ':'.
For example, in \autoref{fig:overview}, we first remove the space in `cookie. name' and then replace the last character `;' with `:'.

\section{Evaluation}

To assess the performance of our system, we measure the Character Error Rate (CER) and Word Error Rate (WER). WER and CER are percentages obtained from the Levenshtein distance between the recognized sequence and the corresponding ground truth. They are calculated as
\[ \frac{D+I+S}{L} \times 100\% \]
where D is the number of deleted units, I is the number of inserted
units, S is the number of substituted units, and L is the total number of
units in the ground truth transcriptions. A unit is a word for WER or a
character for CER.

We evaluate our recognition approach by applying our framework to the 45 code samples in our database. In the following section, we compare the results of our enhanced recognizer to the results of using MyScript alone.

\section{Results} \label{results}

\begin{table}
  \centering
  \begin{tabular}{l r r}
    {}
    & {t-test score ($t_{14}$)}
      & {P-value} \\
    \midrule
    WER on sample 1 & -9.02 & $P < 0.00001$ \\
    WER on sample 2 & -8.29 & $P < 0.00001$ \\
    WER on sample 3 & -6.57 &  $P < 0.00001$  \\
    CER on sample 1 & -3.88  & $P < 0.001$  \\
    CER on sample 2 & -5.45 & $P < 0.0001$ \\
    CER on sample 3 & -6.13 & $P < 0.0001$ \\
  \end{tabular}
  \caption{Statistical evidence (T-test and P-value) for WER and CER on three code samples}~\label{table:significance}
\end{table}

As shown in \autoref{result}, 
our augmented recognition approach results in an 8.6\% word error rate and 3.6\% character error rate, on average, over the three code samples, which outperforms the original MyScript recognizer with 31.31\%  and 9.24\% in word and character error rate respectively. We also find statistical evidence for an effect of our augmented recognition approach on both WER and CER (See \autoref{table:significance}). 


\begin{table}
  \centering
  \begin{tabular}{l r r}
    {System}
    & {WER(\%)}
      & {CER(\%)} \\
    \midrule
    \textcolor{red}{Augmented MyScript}  & \textcolor{red}{8.6} & \textcolor{red}{3.6} \\
    Kozielski et al. \cite{doetsch2013improvements} & 9.5 & 2.7 \\
    Keysers et al. \cite{keysers2016multi} & 10.4 &  4.3  \\
    Zamora et al. \cite{zamora2014neural} & 16.1  & 7.6  \\
    Poznanski et al. \cite{poznanski2016cnn} & 6.45 & 3.44 \\
  \end{tabular}
  \caption{Performance of our system compared to handwritten English recognition systems on the IAM dataset}~\label{table:iam}
\end{table}


Since there is no existing handwriting source code recognizer for comparison, we compare the recognition rate of our our augmented MyScript recognition system (on source code) to that of four state-of-the-art general handwritten English recognition systems (on general text). 
The IAM handwriting database \cite{marti2002iam} consists of 9,285 lines of general handwritten text
written by approximately 400 writers with no restrictions on style or writing tool. This database has been widely used to evaluate English handwriting recognition systems. The four systems in \autoref{table:iam} were tested based on this IAM handwriting database. \autoref{table:significance} shows that the WER and CER of our augmented source code recognition system are comparable with other state-of-the-art handwritten English recognition systems on general handwritten text.




\section{Discussion}


Our approach achieved an 8.6\% word error rate and a 3.6\% character error rate on the collected dataset by taking the language grammar rules into account. Overall, improvement of our recognition pipeline over the baseline MyScript recognition engine can be attributed to addressing the three main error types identified in Section \ref{sec:characterizing}.  After statement concatenation, all unnecessary space errors in a single token are removed. Ensuring the last character of a statement eliminates 32\% of the symbol errors. Token processing fixes around 78\% of the word errors. 

Recognition results, however, are still not 100\% accurate. Initial inspection indicates that this is mainly due to the illegible or cursive handwriting of the participants and the incorrect recognition of symbols. Also, since one of our lexicons is dependent on the non-keywords already recognized in the code, incorrectly recognized words will also be added to the lexicon, thereby corrupting the lexicon and preventing it from enhancing the recognition of the following words. Additionally, it is difficult to identify incorrectly recognized symbols; for example, if `(' appearing in the middle of the text is recognized as `l', it becomes impossible to rectify it using our approach. 
Errors like unmatched `(' and `)' in a statement can be detected, but not reliably corrected. For example, `(name' can be recognized as `cname', but we have no evidence to correct `cname' to `(name'. Two methods can be employed to resolve remaining errors such as this. The first is to develop a widget in the handwriting interface to highlight all errors that are identified but can't be corrected and let users correct them manually. Another option is to train a language model to identify words that do not exist \cite{zamora2014neural}.

Because typing on a virtual keyboard is the standard input method on touchscreen devices, it is useful to examine how virtual keyboard typing error rates compare to those of handwritten source code recognition.
Almusaly et al. report a 7.81\% total error rate (TER) for typing \textit{Java} programs on a standard virtual keyboard as measured from 32 participants \cite{almusaly2015syntax}. TER, similar to CER, is a measure of the total number of errors (i.e., omissions, substitutions, and insertions) and corrections that are made in the resulting typed text.  Our handwriting results are comparable.

This approach can also be generalized to other programming languages with strict grammar rules. For instance, one can define statement classes for \textit{Java} according to the first word in the statement and then replace the regular expressions with productions of \textit{Java} grammar rules.  Algorithms for searching and replacing similar words can be kept unchanged. Other heuristic steps like concatenating tokens are also trivial to implement for new languages. 

\section{Conclusion and Future Work}
The keyboard is not an ideal input mechanism for every person and situation.
Alternatives to typing, such as speech, have been considered in the past \cite{desilets2006voicecode, gordon2013improving}. However, speech is not always an appropriate option given social conventions and privacy issues.
Given advances in pen-based technology that provides an opportunity for users to engage with devices in a potentially more `natural' way than that supported by a virtual keyboard, handwriting input is a viable alternative to virtual keyboard input. In this paper, we have explored handwriting recognition specifically for source code with the ultimate goal of supporting handwriting as a means for programming.
We collect and present a small database of publicly available handwritten source code samples and we propose an approach to recognize handwritten source code by leveraging a commercial handwriting recognition system. Experiments on the data collected from 15 participants show our framework has an average 8.6\% word error rate and 3.6\% character error rate which outperforms the baseline recognition system and produces rates comparable to the recognition of general handwritten English text.  We are encouraged by these initial results but believe there are several avenues of future work.

From the view-point of human-computer interaction, usability and user satisfaction is critical. For handwriting text input, users expect recognition technology with a low error rate and responsive recognition speed. LaLomia et al. \cite{lalomia1994user} reported that users are willing to accept a recognition error rate of only 3\% (a 97\% recognition rate), although Frankish et al. \cite{frankish1995recognition} concluded that users will accept higher error rates depending on the text-editing task. It would not be surprising, therefore, if higher error rates were acceptable for source code entry and editing which is inherently difficult due primarily to the use of symbols. Input speed is another concern with respect to handwriting. Modest touch typing speeds on a virtual keyboard in the range of 20 to 40 words per minute (wpm) are achievable.
Handwriting speeds are commonly in the 15 to 25 wpm range \cite{card1983psychology,devoe1967alternatives,dunlop2009pickup}. We suspect that this decrease in speed, however, will be acceptable to the particular groups for whom handwriting is the most viable input option. Additionally, in professional programming, most of the code that developers
write involves reuse of existing example code and libraries \cite{bellon2007comparison}. This `reuse' typically amounts to editing existing code to suit a
new context or problem and generally provides benefits to developers in terms of time and error reduction \cite{ko2011state}.
For these reasons, we envision our system as being particularly useful in the code editing domain as opposed to writing extensive source code from scratch.  Studying how the algorithms perform in editing tasks is left as future work.



While databases exist for research in general handwritten text recognition \cite{marti2002iam, grosicki1rimes}, there is no such dataset for handwritten source code.  This paper represents the first such contribution of a handwritten source code dataset consisting of 555 lines of  \textit{Python} code written by 15 participants. While we recognize that using the same three code samples for all users and employing a ``copy task'' may lessen the generality of the dataset, we sought to eliminate all effects of cognitive complexity (e.g. actually solving programming problems) to focus solely on the handwritten source code quality.  Collecting data for other programming languages and for actual programming tasks is left as future work.

The next most obvious area of future work is to develop a handwritten source code recognition system from scratch instead of augmenting the results produced by an existing system.  We suspect this approach would lead to comparable and most likely improved recognition rates. Building a universal handwritten source code reading system could employ deep learning techniques such as Concurrent Neural Networks \cite{poznanski2016cnn} or neural network language models \cite{zamora2014neural} trained purely on the source code. 

Additionally, there are several opportunities to explore the integration of handwriting recognition into source code IDEs \cite{frye2008pdp}.  For example, how do we now integrate source code completion into a handwriting-based interaction?   Can we integrate elements such as syntax insertion and highlighting?  Exploring the affordances of handwriting in the context of an IDE is an exciting area of future work that is enabled by these initial findings.



Multimodal methods present another area of future work.  Perhaps the combination of handwriting and speech input or handwriting and occasional keyboard input \cite{mueller2014pen} begin to produce interaction experiences that rival those of typed source code input.  

Finally, we will never reach a perfect recognition rate for handwritten text (general or source code).  How do we effectively support efficient editing of the recognized text so that users can quickly correct mistakes? Natural and effective text entry and editing is an interesting topic for future studies.


\acknowledgments{
The authors wish to thank all the study participants as well as Poorna Talkad Sukumar, Jason Liu, and Suwen Lin for their valuable discussions and input.}

\bibliographystyle{abbrv-doi}

\bibliography{template}
\end{document}